\documentclass[conference]{IEEEtran}
\IEEEoverridecommandlockouts
\usepackage{cite}
\usepackage{amsmath,amssymb,amsfonts}
\usepackage{algorithmic}
\usepackage{graphicx}
\usepackage{caption}
\usepackage{subcaption}
\usepackage{amssymb}
\usepackage{booktabs}
\usepackage{textcomp}
\usepackage{multirow}
\usepackage{array}
\usepackage{hyphenat}
\usepackage{tikz}
\usepackage{algorithm}
\usepackage{algorithmic}
\usepackage{xcolor}
\usepackage{float}
\usepackage{color}
\usepackage[normalem]{ulem}

\usepackage[colorlinks=true, linkcolor=black, urlcolor=black, citecolor=black]{hyperref}
\def\BibTeX{{\rm B\kern-.05em{\sc i\kern-.025em b}\kern-.08em
    T\kern-.1667em\lower.7ex\hbox{E}\kern-.125emX}}

\definecolor{purple}{HTML}{C8A2D0}  
\definecolor{green}{HTML}{A8D5AA}   
\definecolor{blue}{HTML}{A3D5F5}    

\begin{document}

\title{Optimizing Task Scheduling in Fog Computing with Deadline Awareness}

\author{
\IEEEauthorblockN{
Mohammad Sadegh Sirjani,
Mohammad Ahmad,
Amir Mousavi,
Erfan Nourbakhsh,
Khoa Nguyen
}
\IEEEauthorblockA{Department of Computer Science,
University of Texas at San Antonio, San Antonio, USA\\
\{mohammadsadegh.sirjani, mohammad.ahmad, seyedamir.mousavi, erfan.nourbakhsh,khoa.nguyen\}@utsa.edu}
}

\maketitle

\begin{abstract}
The rise of Internet of Things (IoT) devices has led to the development of numerous time-sensitive applications that require quick responses and low latency. Fog computing has emerged as a solution for processing these IoT applications, but it faces challenges such as resource allocation and job scheduling. Therefore, it is crucial to determine how to assign and schedule tasks on Fog nodes. This work aims to schedule tasks in IoT while minimizing the total energy consumption of nodes and enhancing the Quality of Service (QoS) requirements of IoT tasks, taking into account task deadlines. This paper classifies Fog nodes into two categories based on their traffic level: low and high. It schedules short-deadline tasks on low-traffic nodes using an Improved Golden Eagle Optimization (IGEO) algorithm, an enhancement that utilizes genetic operators for discretization. Long-deadline tasks are processed on high-traffic nodes using reinforcement learning (RL). This combined approach is called the Reinforcement Improved Golden Eagle Optimization (RIGEO) algorithm. Experimental results demonstrate that RIGEO achieves up to a 29\% reduction in energy consumption, up to an 86\% improvement in response time, and up to a 19\% reduction in deadline violations compared to state-of-the-art algorithms.
\end{abstract}
\section{Introduction}
In recent years, the Internet of Things (IoT) has become a prominent technology in the internet and networking industry. It consists of a system of linked physical devices, like household items and cars, that have sensors, software, and online connectivity \cite{nematiollahi2021offloading}. Although IoT has been expanding quickly, it is encountering obstacles such as traffic congestion, delays, and excessive energy usage.

Adhering to deadlines is crucial in real-time IoT applications such as: e-health in healthcare, smart grid in energy management, livestock monitoring in agriculture, and smart traffic in transportation management. These time-sensitive applications have stringent QoS requirements including response time, deadline adherence, and energy efficiency that exceed the capabilities of resource-constrained IoT devices \cite{ghaffari2025qte}.

Fog computing addresses these challenges by providing computational capabilities at the edge, reducing delays and energy consumption. However, fog computing's diverse and limited resources necessitate efficient scheduling and allocation strategies \cite{azami2023deadline}. Many existing solutions rely on single optimization strategies or fail to account for the dynamic, heterogeneous nature of IoT-Fog environments \cite{ghanavati2023dynamic}.

\begin{figure}[t!]
    \centering
    \includegraphics[width=1.0\columnwidth, height=5.5cm, keepaspectratio]{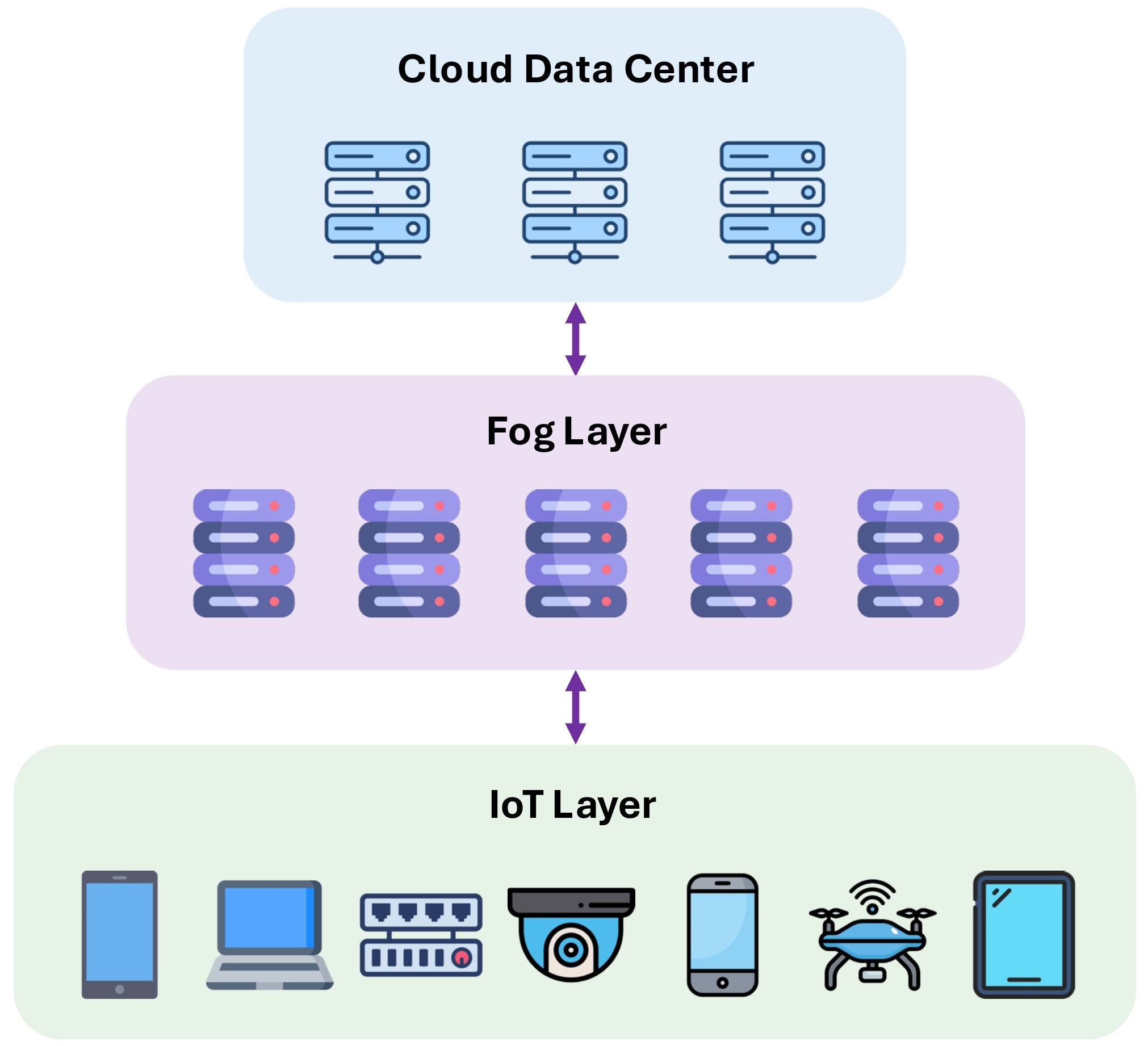}
    \caption{The architecture of the IoT-Fog-Cloud network.}
    \label{fig:fig-1}
\end{figure}

Meta-heuristic algorithms are particularly valued for their speed and ability to quickly find satisfactory solutions in complex optimization problems. Among these approaches, the Golden Eagle Optimization (GEO) algorithm, inspired by the hunting behavior of golden eagles, demonstrates effectiveness in global optimization tasks. Its computational efficiency is especially crucial for preventing deadline violations in time-sensitive IoT applications.

This work proposes the Reinforcement Improved Golden Eagle Optimization (RIGEO) algorithm, an adaptive task scheduling framework. RIGEO dynamically classifies Fog Nodes (FNs) into low-traffic and high-traffic categories based on real-time network conditions. Low-traffic nodes are characterized by average network traffic that falls below a predefined threshold, whereas high-traffic nodes exceed this average. It employs Improved Golden Eagle Optimization (IGEO) for short-deadline tasks on low-traffic nodes and Reinforcement Learning for long-deadline tasks on high-traffic nodes, leveraging the strengths of each method for different operational scenarios \cite{sirjani2025controller}.

RIGEO operates within a three-layered architecture: cloud, fog, and IoT devices. Figure~\ref{fig:fig-1} illustrates this hierarchy, where the fog layer comprises a controller and a network for resource management and task scheduling, while the cloud offers significant computing and storage capabilities. 

Task scheduling faces multiple challenges, including energy efficiency, meeting deadline constraints, reducing latency, and ensuring reliability. The following subsections review relevant studies in these areas.

\subsection{Energy Efficiency}
Wang et al.~\cite{wang2018emef} propose MEETS, a task scheduling algorithm for fog computing that determines optimal task allocation to fog nodes before scheduling tasks on each node. MEETS significantly reduces energy consumption while meeting all task deadlines; however, it assumes homogeneous fog networks and constant task arrival rates, which may not be realistic in practice.

Vispute et al.~\cite{vispute2023energy} propose a PSO-based task scheduling scheme that outperforms existing approaches in makespan, energy consumption, and execution time.

\subsection{Deadline Constraints}
Azami et al.~\cite{azami2023deadline} incorporate two semi-greedy algorithms to minimize deadline violations for IoT tasks while ensuring QoS compliance, achieving a significant reduction in deadline violation time.

\subsection{Latency Reduction}
Klatoun et al.~\cite{klatoun2022novel} propose a task scheduling framework for IoMT that processes data closer to end-users, reducing latency and enhancing response times.

Hosseini et al.~\cite{hosseini2022optimized} present an algorithm using fuzzy logic and AHP to prioritize tasks based on completion time, energy, RAM, and deadlines, effectively balancing cost and latency in mobile fog environments.


\begin{table}[t]
    \centering
    \footnotesize
    \begin{tabular*}{\columnwidth}{@{\extracolsep{\fill}}ccccc}
        \toprule
        \multirow{3}{*}{\textbf{Ref.}} & \multirow{3}{*}{\textbf{Year}} & \multicolumn{3}{c}{\textbf{Factors}} \\
        \cmidrule(lr){3-5}
        & & \textbf{Energy} & \textbf{Deadline} & \textbf{Latency} \\
        & & \textbf{Efficiency} & \textbf{Constraints} & \textbf{Reduction} \\
        \midrule
        \cite{wang2018emef} & 2018 & \checkmark & \checkmark & \texttimes \\
        \cite{vispute2023energy} & 2023 & \checkmark & \texttimes & \texttimes \\
        \cite{azami2023deadline} & 2022 & \checkmark & \checkmark & \texttimes \\
        \cite{klatoun2022novel} & 2022 & \checkmark & \checkmark & \checkmark \\
        \cite{hosseini2022optimized} & 2022 & \checkmark & \checkmark & \checkmark \\
        \cite{ghanavati2023dynamic} & 2020 & \texttimes & \texttimes & \texttimes \\
        \midrule
        \textbf{Ours} & \the\year & \checkmark & \checkmark & \checkmark \\
        \bottomrule
    \end{tabular*}
    \caption{Comparison of factors considered in related work}
    \label{tab:comparison}
\end{table}

Table~\ref{tab:comparison} presents an overview of these works. This paper optimizes system response time, deadline violation time, and energy consumption in FNs. The key contributions include:
\begin{itemize}
    \item We propose the IGEO algorithm for the efficient scheduling of short-deadline tasks.
    \item We introduce a Reinforcement Learning-based approach for the scheduling of tasks with long deadline constraints.
    \item We conduct comprehensive experiments to evaluate the performance of our proposed framework in terms of energy consumption, deadline violation rates, and response time metrics within fog computing environments.
\end{itemize}


\section{Methodology}
\label{sec:methodology}
This section outlines the Task Scheduling problem in mathematical terms and provides a detailed description of the proposed method.

\subsection{Problem Statement}
\label{subsec:problem-statement}
Given a set of $n$ independent tasks $T = \{t_1, t_2, \ldots, t_n\}$ that must be assigned to $m$ heterogeneous FNs $F = \{f_1, f_2, \ldots, f_m\}$, the goal is to assign tasks to nodes in a way that minimizes energy consumption and ensures that the task deadlines are met. The fog network is represented as a graph $G = (V, E)$ with connection links between nodes. Each task $t_i$ has a deadline $d_i$ in milliseconds.

Each task $t_i$ is described by a tuple $(c_i, m_i, d_i, a_i, q_i)$ sampled as $c_i \sim \mathcal{U}\{1,\ldots,5\}$, $m_i,d_i,q_i \sim \mathcal{U}[0,1]$, and $a_i \sim \mathcal{U}\{1,\ldots,T\}$ where $T=10$, so batch sizes vary naturally across time slots. Each task also carries an implicit data payload whose size is proportional to $c_i$ (typically $s_i \approx 0.1$--$2.5$~MB for IoT sensor and actuation data).

Each FN $f_j$ is described by a tuple $(C_j, M_j, \kappa_j, \theta_j, \varepsilon_j, h_j)$ where $C_j \sim \mathcal{U}\{81,\ldots,100\}$, $M_j \sim \mathcal{U}\{5,\ldots,16\}$~GB, $\kappa_j \sim \mathcal{U}[0,1]$, $\theta_j \sim \mathcal{U}[0.5,1.5]$, $\varepsilon_j = u_j + v_j$ with $u_j \sim \mathcal{U}\{3,4,5\}$ and $v_j \sim \mathcal{U}[0,1]$, and fixed $h_j = 0.003$.

The fog network is modeled as a fully connected graph $G=(V,E)$ with $|V|=m$ nodes on high-bandwidth wired links ($B_{jk} \geq 1$~Gbps, $\delta_{jk} < 0.1$~ms), making communication overhead negligible relative to execution time (see Section~\ref{subsubsec:response-time}). A symmetric communication-preference matrix $P \in [0,1]^{m \times m}$ with $P(i,i)=0$ and $P(i,j)=P(j,i) \sim \mathcal{U}[0,1]$ is refined online by the RL stage.

The parameters used in the methodology are defined as follows:

\subsubsection{Response time} \label{subsubsec:response-time}
The response time $R_i$ for task $t_i$ assigned to FN $f_j$ is~\cite{nematiollahi2021offloading}:
\begin{equation}
R_i = P_i + T_i + E_i
\label{eq:response_time}
\end{equation}

where the three components are defined as follows.

\textit{Propagation delay} -- the signal travel time over the physical link between the source and the assigned FN:
\begin{equation}
P_i = \delta_{\text{src},\, j}
\label{eq:prop_delay}
\end{equation}

\textit{Transmission delay} -- the time to transfer the task's data payload $s_i$ over the link bandwidth:
\begin{equation}
T_i = \frac{s_i}{B_{\text{src},\, j}}
\label{eq:trans_delay}
\end{equation}

\textit{Execution time} -- the CPU processing time on the assigned node:
\begin{equation}
E_i = \frac{c_i}{C_j + 1}
\label{eq:exec_time}
\end{equation}

Since $P_i + T_i \ll E_i$, the per-task response time simplifies to $R_i \approx E_i$, and the system-wide metric is the makespan $\max_j W_j$, where $W_j = \sum_{x_k=j} E_k$. Deadline violations are thus:
\begin{equation}
DV_i = \max\!\bigl(0,\; E_i - d_i\bigr)
\label{eq:dv_approx}
\end{equation}

\subsubsection{Deadline violation}
\label{subsubsec:deadline-violation}
A deadline violation occurs when a task's completion time exceeds its specified deadline~\cite{ghaffari2024deadline}.

The deadline violation time for each task is calculated using Eq.~(\ref{eq:deadline_violation_task}), and the total violation time is determined using Eq.~(\ref{eq:deadline_violation_total}).

\begin{equation}
    DV_i = \max(0, R_i - d_i)
    \label{eq:deadline_violation_task}
\end{equation}

\begin{equation}
    DV_{total} = \sum_{i=1}^{n} DV_i
    \label{eq:deadline_violation_total}
\end{equation}

\subsubsection{Energy Consumption}
\label{subsubsec:energy-consupmtion}
The energy used by FNs to complete all tasks is calculated by adding the energy consumed during active ($E_{active}$) and idle modes ($E_{idle}$). The coefficients $\alpha$ and $\beta$ are user-determined parameters.

Eq.~(\ref{eq:energy_fn}) shows energy consumption by $f_j$ and Eq.~(\ref{eq:energy_total}) shows total system energy consumption~\cite{ghaffari2024deadline}.

\begin{equation}
    E_j = \alpha \cdot E_{active,j} + \beta \cdot E_{idle,j}
    \label{eq:energy_fn}
\end{equation}

\begin{equation}
    E_{total} = \sum_{j=1}^{m} E_j
    \label{eq:energy_total}
\end{equation}

\subsection{Simulation Environment}
\label{subsec:simulation}



We evaluate $n \in \{200,300,400,500,600\}$ tasks on $m=20$ heterogeneous FNs over $T=10$ time slots (Table~\ref{tab:sim-params}), averaged over 50 Monte Carlo runs. FNs whose mean traffic exceeds the network-wide median are labeled high-traffic ($F_{\text{high}}$); the deadline threshold is the 25th percentile of each workload. Population-based optimizers use $N_{pop}=50$ and $N_{iter}=2$ per batch.

\begin{table}[t]
    \centering
    \footnotesize
    \begin{tabular*}{\columnwidth}{@{\extracolsep{\fill}}ll}
        \toprule
        \textbf{Parameter} & \textbf{Value} \\
        \midrule
        Fog nodes ($m$) & 20 \\
        Network topology & Full mesh \\
        Task loads evaluated & 200, 300, 400, 500, 600 \\
        Time slots ($T$) & 10 \\
        Monte Carlo runs & 50 \\
        Population size ($N_{pop}$) & 50 \\
        Iterations per batch ($N_{iter}$) & 2 \\
        Traffic history length & 200 steps \\
        Max node traffic load & 100 units \\
        \midrule
        FN CPU capacity ($C_j$) & $\mathcal{U}\{81,\ldots,100\}$ units \\
        FN memory ($M_j$) & $\mathcal{U}\{5,\ldots,16\}$~GB \\
        FN energy rate ($\varepsilon_j$) & $\mathcal{U}\{3,4,5\}+\mathcal{U}[0,1)$~J/cycle \\
        FN energy harvest ($h_j$) & $0.003$ \\
        FN cost rate ($\kappa_j$) & $\mathcal{U}[0,1)$ \\
        FN QoS throughput ($\theta_j$) & $\mathcal{U}[0.5,1.5]$ \\
        \midrule
        Task data size ($s_i$) & $\leq c_i \times 0.5 \approx 2.5$~MB \\
        Link bandwidth ($B_{jk}$) & $\geq 1$~Gbps (wired LAN) \\
        Propagation delay ($\delta_{jk}$) & $< 0.1$~ms (co-located) \\
        Communication overhead ($P_i + T_i$) & Negligible ($\ll E_i$); see Sec.~\ref{subsubsec:response-time} \\
        \midrule
        Deadline threshold & 25th percentile \\
        Traffic threshold & Median of node mean \\
        \bottomrule
    \end{tabular*}
    \caption{Simulation parameters}
    \label{tab:sim-params}
\end{table}

\subsection{Proposed Method}
At the start of each experiment, traffic labels and the deadline threshold are computed and used to partition FNs and tasks.

When network traffic is high, faster algorithms must be used to ensure that tasks are executed within their deadlines.

The proposed RIGEO algorithm considers deadline tasks in two stages:
\begin{enumerate}
    \item Short-deadline tasks execute on low-traffic FNs to prevent queue waiting.
    \item Long-deadline tasks are processed using Reinforcement Learning on high-traffic FNs, as meta-heuristic overhead isn't justified for tasks with more relaxed timing constraints.
\end{enumerate}

\subsubsection{Reinforcement Learning for Long-Deadline Tasks}

RIGEO employs a table-based RL policy to schedule long-deadline tasks on $F_{\text{high}}$. The agent maintains the communication-preference matrix $P \in [0,1]^{m \times m}$ (initialized as described in Section~\ref{subsec:simulation}), where $P(s,f)$ encodes the learned preference for assigning a task to FN $f$ given that the preceding task was assigned to FN $s$. The matrix is updated online as each task in the batch is processed.

\textbf{State.} $s \in F_{\text{high}}$: the index of the most recently assigned high-traffic FN. The initial state is drawn uniformly at random from $F_{\text{high}}$.

\textbf{Action.} Given state $s$, the agent greedily selects the highest-valued reachable FN:
\begin{equation}
    a = \underset{f \in F_{\text{high}}}{\arg\max}\; P(s, f).
    \label{eq:rl_action}
\end{equation}

\textbf{Reward and penalty.} After task $t_i$ is assigned to FN $a$, the response time $R_i \approx E_i = c_i / (C_a + 1)$ (see Section~\ref{subsubsec:response-time}) is compared with deadline $d_i$. A met deadline rewards the transition (scales up by 10\%); a violation penalises it (scales down by 10\%):
\begin{equation}
    P(s,a) \leftarrow
    \begin{cases}
        \min\!\bigl(1,\; P(s,a) \times 1.1\bigr) & \text{if } R_i \leq d_i, \\[2pt]
        \max\!\bigl(0,\; P(s,a) \times 0.9\bigr) & \text{if } R_i  >  d_i.
    \end{cases}
    \label{eq:rl_update}
\end{equation}

\textbf{State transition.} The state advances to the assigned node, $s \leftarrow a$, so the policy adapts sequentially as tasks within the batch arrive.

\subsubsection{Golden Eagle Optimization (GEO)}
GEO is a swarm-based meta-heuristic whose search mechanism balances exploration (searching new areas) and exploitation (refining solutions). It adjusts speed during different hunting stages to find global optima while avoiding local optima.

In GEO, each eagle selects a target, calculates its cruising vector, and circles the best location found. The prey represents the optimal solution, with each eagle tracking its own best discovery. The attack vector for eagle $i$ is calculated through Eq. (\ref{eq:attack}), where $\vec{X}_f^*$ is the position vector of the best prey found so far and $\vec{X}_i$ is the current position vector of eagle $i$.

\begin{equation}
\vec{A}_i = \vec{X}_f^* - \vec{X}_i \label{eq:attack}
\end{equation}

The step vector for eagle $i$ at iteration $t$ is given by Eq. (\ref{eq:step}). Random vectors $\vec{r}_1$ and $\vec{r}_2$ have elements in $[0,1]$, with user-defined coefficients $p_a$ (attack) and $p_c$ (cruise). Position updates follow Eq. (\ref{eq:position}).

\begin{equation}
\Delta x_i = \vec{r}_1 p_a \frac{\vec{A}_i}{\|\vec{A}_i\|} + \vec{r}_2 p_c \frac{\vec{C}_i}{\|\vec{C}_i\|} \label{eq:step}
\end{equation}

\begin{equation}
x^{(t+1)} = x^t + \Delta x_i^t \label{eq:position}
\end{equation}

\subsubsection{Improved Golden Eagle Optimization (IGEO)}
Since task scheduling is discrete, GEO must be discretized using genetic operators that enhance both exploration and exploitation capabilities, enabling the discovery of optimal solutions in a minimal amount of time. IGEO enhances GEO through genetic operators for discrete task assignment \cite{mohammadi2021golden}.

Unit vectors indicate direction without size information. Vector sizes depend only on $\vec{r}_1 p_a$ and $\vec{r}_2 p_c$. When $|r_1 p_a| < |r_2 p_c|$, the cruise tendency is higher, employing mutation for a high search power. When $|r_1 p_a| > |r_2 p_c|$, the attack tendency is higher, using crossover operators.

\begin{figure}[t!]
    \centering
    \includegraphics[width=0.7\columnwidth, height=3.5cm, keepaspectratio]{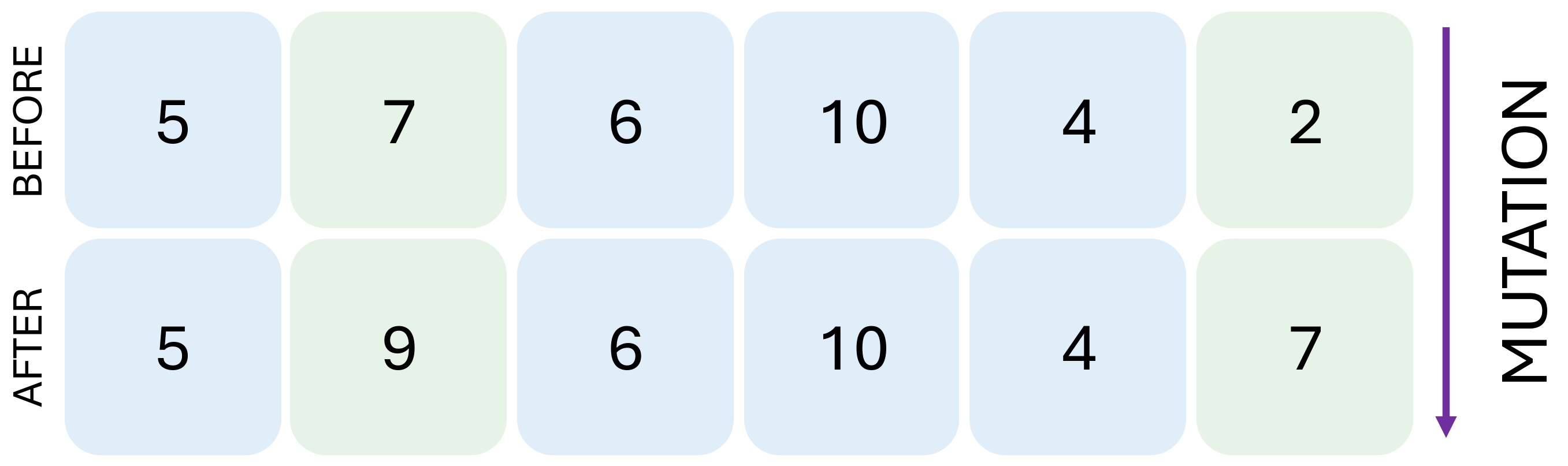}
    \caption{Mutation operator. \textcolor{green}{Green} segments indicate genes subject to mutation, \textcolor{blue}{Blue} segments remain constant.}
    \label{fig:mutation}
\end{figure}

Negative step vectors in Eq. (\ref{eq:step}) indicate exploration tendency; positive vectors indicate exploitation. IGEO employs different genetic operators accordingly. For negative step vectors, mutation replaces addition in Eq. (\ref{eq:position}). Figure~\ref{fig:mutation} illustrates the mutation operator, which randomly alters the parental genetic material.

Eq. (\ref{eq:igeo_negative}) shows the IGEO operation for negative step vectors, where $x_{best}$ is the best found location and $x_t$ is the current iteration location.

\begin{equation}
x_{t+1} = \begin{cases}
\text{Mutation}(x_{best}) & \text{if } r \geq 0.5 \\
\text{Mutation}(x_t) & \text{if } r < 0.5
\end{cases} \label{eq:igeo_negative}
\end{equation}

\begin{figure}[t!]
    \centering
    \includegraphics[width=1.0\columnwidth, height=5cm, keepaspectratio]{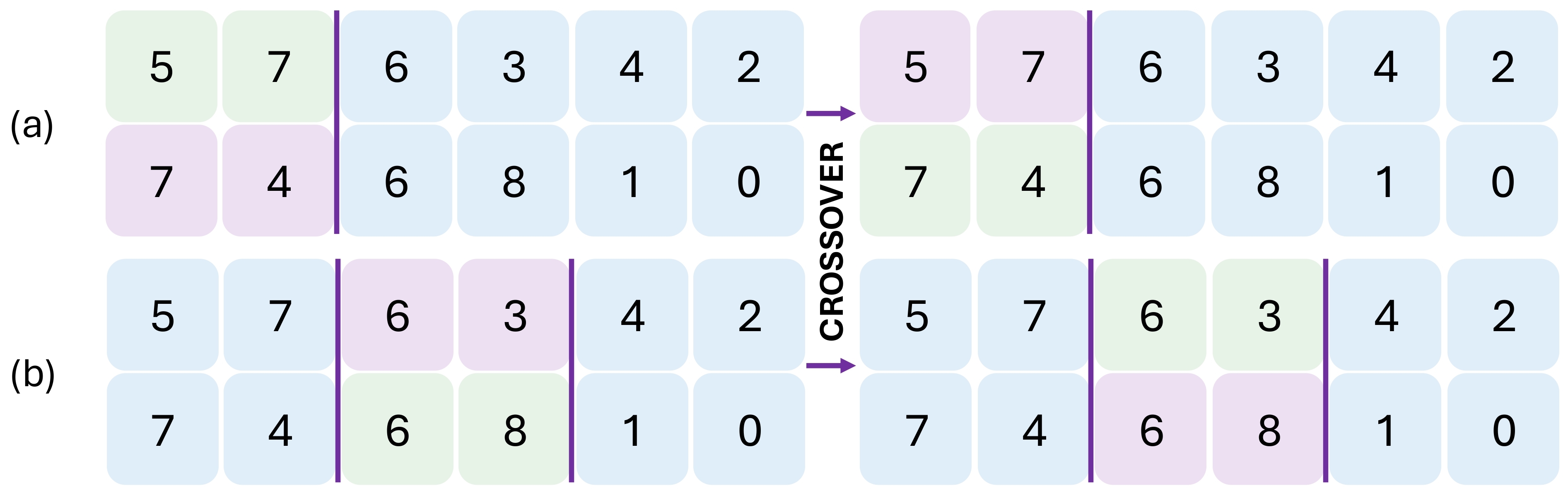}
    \caption{Crossover operations: (a) one-point and (b) two-point. \textcolor{purple}{Purple} and \textcolor{green}{Green} indicate exchanged segments.}
    \label{fig:crossover}
\end{figure}

For positive step vectors, crossover operators provide efficiency. Figure~\ref{fig:crossover} illustrates how genes from two parents exchange at predetermined cutting points. Eq. (\ref{eq:igeo_positive}) shows the update for positive operations.

\begin{equation}
x_{t+1} = \begin{cases}
\text{single-point crossover}(x_{best}, x_t) & \text{if } r \geq 0.5 \\
\text{two-point crossover}(x_{best}, x_t) & \text{if } r < 0.5
\end{cases} \label{eq:igeo_positive}
\end{equation}

Algorithm~\ref{alg:rigeo} shows how short-deadline tasks are sent to low-traffic FNs for IGEO processing, while long-deadline tasks go to high-traffic FNs for RL processing.

\begin{algorithm}[t!]
    \caption{RIGEO Task Scheduling Algorithm}
    \label{alg:rigeo}
    \footnotesize
    \begin{algorithmic}[1]
        \REQUIRE Task $T$ with deadline $D$, Fog nodes $F = \{f_1, \ldots, f_n\}$
        \ENSURE Scheduled and processed task
        \STATE $Low\_Traffic\_FNs \gets \emptyset$, $High\_Traffic\_FNs \gets \emptyset$
        \FOR{each $f \in F$}
            \IF{$f.traffic < traffic\_threshold$}
                \STATE Add $f$ to $Low\_Traffic\_FNs$
            \ELSE
                \STATE Add $f$ to $High\_Traffic\_FNs$
            \ENDIF
        \ENDFOR
        \IF{$T.deadline < deadline\_threshold$}
            \STATE $f \gets SELECT(Low\_Traffic\_FNs)$
            \STATE $result \gets IGEO\_PROCESS(T, f)$
        \ELSE
            \STATE $f \gets SELECT(High\_Traffic\_FNs)$
            \STATE $result \gets RL\_PROCESS(T, f)$
        \ENDIF
        \RETURN $result$
    \end{algorithmic}
\end{algorithm}

\begin{figure*}[!t]
    \centering
    \includegraphics[width=\textwidth]{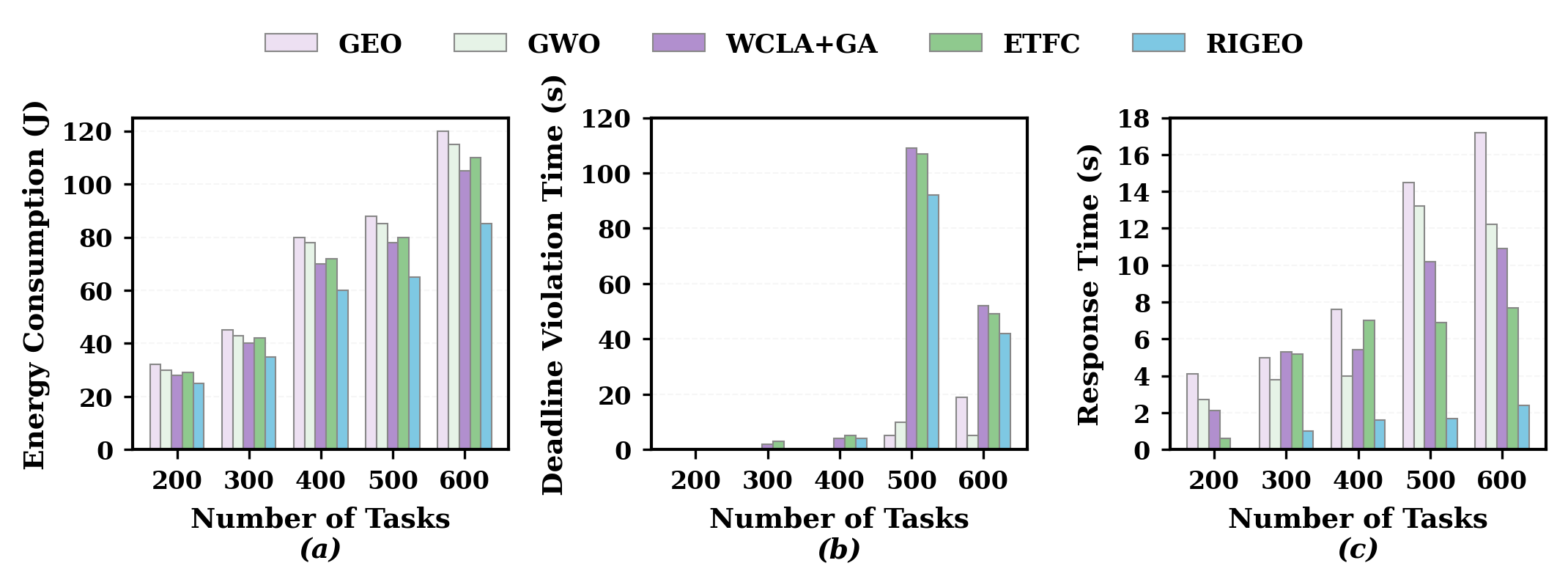}
    \caption{Performance metrics with varying task numbers (200-600) and 20 FNs over 50 runs: (a) Energy Consumption, (b) Deadline Violation, (c) Response Time}
    \label{fig:performance-metrics}
\end{figure*}

\section{Results}
\label{sec:results}
This section presents an appraisal of the RIGEO approach's performance through various metrics. Algorithms were implemented in MATLAB 2016 on an Intel Core i7 2.80 GHz machine with 16 GB RAM.

Results are compared against GEO \cite{mohammadi2021golden}, GWO \cite{banes2022cloud}, WCLA+GA \cite{azami2023deadline}, and ETFC \cite{vispute2023energy} algorithms to showcase RIGEO's performance. Each experimental configuration was executed 50 times with different random task distributions and node initializations. The results presented represent average values across these runs.

Simulation parameters are detailed in Table~\ref{tab:sim-params}.

Experimental results demonstrate that RIGEO reduces response time, minimizes deadline violation time, and optimizes energy consumption compared to other algorithms. Figure~\ref{fig:performance-metrics} shows the performance metrics across varying task numbers. The following sections examine RIGEO's performance on each parameter.

\subsection{Energy Consumption}
Figure~\ref{fig:performance-metrics}(a) illustrates the energy usage of FNs with varying task numbers (200 to 600) and 20 FNs. Energy consumption is calculated using Eq.~(\ref{eq:energy_total}).

RIGEO optimizes FN energy consumption by effectively distributing tasks among nodes, reducing overall usage. At maximum load (600 tasks), RIGEO achieved approximately 29\% lower energy consumption than GEO and 26\% lower than GWO. Comparison with other scheduling techniques reveals RIGEO's consistent success in minimizing energy consumption across all task loads.

\subsection{Deadline Violation}
RIGEO improves deadline adherence compared to other methods, as indicated by the decrease in violation times shown in Figure~\ref{fig:performance-metrics}(b). The method consistently selects nodes with the shortest violation times, even in the presence of node shortages, demonstrating flexibility in addressing deadline breaches.

At 600 tasks, RIGEO demonstrated approximately 19\% improvement over WCLA+GA and 14\% improvement over ETFC in terms of deadline violation time.

\subsection{Response Time}
Response time, the period required for the system response to a task, was systematically integrated into the fitness function as a critical parameter. Figure~\ref{fig:performance-metrics}(c) illustrates RIGEO's superior performance in minimizing response time compared to other approaches with 20 FNs.

RIGEO achieved approximately 86\% faster response time than GEO and 80\% faster than GWO at maximum load. Response time computation is based on Eq.~(\ref{eq:response_time}). The results demonstrate RIGEO's effectiveness in significantly reducing system latency across all task loads.

\section{Conclusion}
\label{sec:conclusion}

This work proposes the Reinforcement Improved Golden Eagle Optimization algorithm to address task scheduling challenges in IoT-fog computing environments. RIGEO dynamically classifies networks into low and high traffic categories, applying IGEO for low-traffic scenarios and Reinforcement Learning for high-traffic conditions. 

Experimental results demonstrate that RIGEO achieves up to a 29\% reduction in energy consumption, up to an 86\% faster response time, and up to 19\% fewer deadline violations compared to state-of-the-art algorithms. The proposed framework successfully balances computational efficiency and scheduling performance through the adaptive selection of algorithms. Future work will explore statistical validation and adaptive threshold optimization.

\bibliographystyle{ieeetr}
\bibliography{references}

\end{document}